# Machine learning based in situ quality estimation by molten pool condition-quality relations modeling using experimental data

**Authors**

Noopur Jamnikar[a,b], Sen Liu[a,b+], Craig Brice[a,b], and Xiaoli Zhang[a,b*]
[*]Corresponding authors: xlzhang@mines.edu

**Affiliations**

[a] Mechanical Engineering, Colorado School of Mines, Golden, CO 80401 USA
[b] The Alliance for the Development of Additive Processing Technologies, Colorado School of Mines, Golden, CO 80401 USA
[+]Co-author

**ABSTRACT:** The advancement of machine learning promises the ability to accelerate the adoption of new processes and property designs for metal additive manufacturing. The molten pool geometry and molten pool temperature are the significant indicators for the final part's geometric shape and microstructural properties for the Wire-feed laser direct energy deposition process. Thus, the molten pool condition-property relations are of preliminary importance for in situ quality assurance. To enable in situ quality monitoring of bead geometry and characterization properties, we need to continuously monitor the sensor's data for molten pool dimensions and temperature for the Wire-feed laser additive manufacturing (WLAM) system. We first develop a machine learning convolutional neural network (CNN) model for establishing the correlations from the measurable molten pool image and temperature data directly to the geometric shape and microstructural properties. The multi-modality network receives both the camera image and temperature measurement as inputs, yielding the corresponding characterization properties of the final build part (e.g., fusion zone depth, alpha lath thickness). The performance of the CNN model is compared with the regression model as a baseline. The developed models enable molten pool condition-quality relations mapping for building quantitative and collaborative in situ quality estimation and assurance framework.

## 1. Introduction

Additive manufacturing is gaining wide adoption in the data-driven modeling industry for building customized products. Low volume of production, high manufacturing cost, and customized products make data-driven modeling for quality assurance and control challenging. Data-driven modeling based upon the experimental data is a step towards real-time monitoring and control of geometric and microstructural properties, in contrast, to post-process characterization. Metallic AM processes are classified into powder bed fusion, direct energy deposition, or sheet lamination. This paper focuses on Wire feed laser additive manufacturing (WLAM), which is a direct energy deposition process. In this method, a focused energy source and deposition material are concentrated at a focal point, commonly in the presence of inert gas. Energy is guided to a narrowly focused area, mainly powered by a source such as a laser beam, electron beam, plasma arc, etc., to melt the deposited materials along with the substrate [1, 2]. The material generally used is either in powder or wire form. The material is deposited in a layer by layer form until the part reaches a near-net shape. In this paper, we focus on the laser power source concentrated on the substrate to melt the heated wire from the wire feeder in the presence of argon gas.

In situ measurement of final build quality is a trending topic in the metal AM industry, allowing a high volume of production and low post-processing cost. Post-processing characterization parameters are crucial for quality and microstructural properties obtained from the build. The output build parameters cannot be measured directly during the process, but knowing the final build properties is vital for quality indication. Most of the research involving build quality analysis is focused on final build geometry and microstructural properties. It involves modeling of surface roughness [3, 4], modeling of microstructural and mechanical properties [5-12], topography, and dimensional accuracy modeling [5, 6, 13-16], either using analytical or analytical-numerical techniques. Other modeling efforts that have been conducted are for flow phenomena and droplet kinematics [10, 17-22] and heat transfer [6, 16, 22]. Final bead geometry modeling and estimation work are mainly developed using thermal history, energy input from the process parameters, or data-driven techniques. For example, Baturynska et al. [23] compared the linear and non-linear regression model's performance for predicting the length, width, and thickness of the build. The input to the model is the part orientation, STL properties, and part placement, and concluded that is obtained using the linear regression for predicting width. While modeling using a machine learning approach and sensing data allows estimating the final build geometry directly while the printing is in progress. For example, Xiong et al. [24] presented a passive two camera vision system for real-time prediction of bead height and width. Image processing and filtering techniques were applied to the camera images, and the experiments were used to validate the method for the GMAW setup. Lu et al. [25] predicted the reinforcement and penetration depth of the weld using image data and optimized Resnet34 as the network structure. Goncalves et al. [26] developed six different CNN architectures to analyze the molten pool images to yield the height and width of the clad bead as outputs. Wang et al. [27] developed a regression model for predicting the bead geometric parameters, i.e., height, width, accumulated area, penetration, penetration area, dilution ratio, and aspect ratio using process parameters.

Microstructural properties modeling has been done widely using finite element analysis (FEA), thermal modeling, and simulation based approach. Many research studies have been conducted to analyze the effect of system and thermal parameters on the microstructural data. Recent work in microstructural properties is focused on the usage of sensing data for microstructural parameter prediction. Farshidianfar et al. [28] developed a real-time microstructure controller using the infrared thermal imaging system and feedback PID control algorithm. The microstructure was controlled by keeping the cooling rate and molten pool temperature within the desired set range. Khanzadeh et al. [29] acquired molten pool thermal images using a dual-wavelength pyrometer to identify patterns using a supervised machine learning technique. A morphological model was developed using processed images for obtaining microstructural information for the build. Imani et al. [30] presented a novel DNN model to learn and classify flaws from a layerwise thermal imaging profile. The developed model is shown to have an accuracy of 92.50% for detecting flaws using fine-grained information. Mahmoudi et al. [31] worked on a layerwise anomaly detection system using the thermal images of the molten pool temperature for a laser powder bed fusion. A Gaussian model is developed to identify the spatial patterns, and a trained classifier determines if the process is in or out of control. Zhang et al. [32] presented a CNN structure to classify the occurrence of porosity in direct laser deposition with 91.2% accuracy from image data. Gonzalez et al. [33] presented a novel convolutional laser based manufacturing (ConvLBM) method to extract features from images for the laser-based manufacturing process in real-time and estimate the printed bead dilution and location of defects. Bartlett et al. [34] developed a microstructural defect classification model using the 3D digital image correlation imaging system. A Naïve-Bayes classification technique is used for predicting the developed physical defects.

Measurement of post-process characterization data for additive manufactured parts is costly and time-consuming. The existing methods for post-processing parameters measurement involve destructible testing and highly complex systems. The destructible methods consume not only time but also energy and do irreversible damage to the build. Hence, in situ characterization of the final build part is an urgent need and requires the use of data-driven modeling techniques and real-time sensing data. As seen from the literature, bead geometry prediction using sensing data is focused only on a couple of parameters, including bead height and penetration depth. Most of the data-driven characterization modeling involves a classification problem for microstructure. Few researchers have presented an in situ quality assurance architecture predicting the geometric and microstructural properties for the WLAM process using ML techniques. Estimating characterization properties directly from real-time sensing data has rarely been done as it requires a good quality dataset to develop an in situ framework. High manufacturing cost and extremely difficult and costly post-processing characterization impact building a high volume of datasets for data-driven modeling. The build geometry estimation is relatively easy, but the cost of microstructural data estimation is high, time-consuming, and requires highly sophisticated machinery.

To solve this issue, we build an experimental dataset including both geometric and microstructural data for the Ti-6Al-4V and the corresponding sensor and process parameter data under a set of controlled experiments. Thus, utilizing the characterization knowledge gained from experimental data will help estimate and control the print quality properties. Machine learning will serve as an excellent tool to maps the complex relationship between the measurable, controllable, and quality parameters. The Sensing- Quality (S-Q) model development allows a real-time quality assurance framework with improved performance accuracy. The contributions for building an in situ quality monitoring framework include: (a) Experimental data collection for different settings of process parameters for building the final build quality database, including the real-time sensing data. (b) Measurement of four final build geometric parameters, i.e., bead height (H), bead width (W), fusion zone depth (D), and fusion zone area (A) for experimental deposits. (c) Microstructural properties for single bead deposition, i.e., alpha lath thickness ($\alpha$), beta grain length parallel to the build direction ($\beta\|$), and beta grain length perpendicular to the build($\beta\perp$) used in model development. (d) Developed multi-modality CNN models for in situ prediction of characterization properties of the final build part using analyzed real-time monitoring data

## 2. Materials and methodology

*2.1. WLAM experimental setup*

The WLAM DED system has been developed and installed at Oak Ridge National Laboratory in Knoxville, Tennessee. A 6kW laser is delivered to the end effector of the robot arm in the presence of argon filled environment. The feedstock is 1.5875 mm Ti-6Al-4V welding wire per AMS 4954K specification. The laser WLAM DED robot setup, along with the mounted sensors, is shown in Figure 1. The sensors were selected to capture as much data as possible from the process during operation. There were five categories of data that were collected: (1) visual, (2) thermal, (3) positional, (4) chemical, and (5) acoustic. Two Prosilica GT1930C cameras were mounted to the robot head, one directly coaxial with the process and the other at a 90 degree oblique angle to the primary direction of travel. The camera is connected using an Ethernet interface, recording 1936 × 1216 pixels images at 25 frames per second (fps) using the NI PXIe-8234 vision module. The main issue in the laser-based AM process is that the image contrast for the molten pool is too bright to capture the surface morphology directly. Hence, bandpass filters are mounted in front of the camera to reduce the intensity. There are three pyrometers with a temperature range of 50-400˚C, 200-1500˚C, and 1000-2000˚C to measure the leading (Optris CTlaser 3M), trailing (Optris CT XL 3M), and molten pool (Optris CTlaser 05M) temperatures, respectively. The pyrometers are calibrated for emissivity using a heated plate and physical contact measurements with thermocouples. Note that the molten pool pyrometer could not be easily calibrated using a similar method, so the presented data are considered relative and not absolute. The leading and trailing edge pyrometers are pointed approximately 25 mm in front of and behind the molten pool. Temperature data is collected at 100 Hz using NI PXIe-4302 analog input module. An acoustic sensor is mounted on the laser head operating a frequency of 1 kHz to capture variations during the build. Analysis of the sensor signals relative to the process during stable and unstable operation will determine how each sensor can be used in the control logic. The National Instrument (NI) industrial controller NI PXIe-8880 along with the vision development module and analog/digital I/O module, is used for monitoring and controlling the laser DED system. This paper focuses on studying the in situ sensing data from the coaxial camera and three pyrometers.

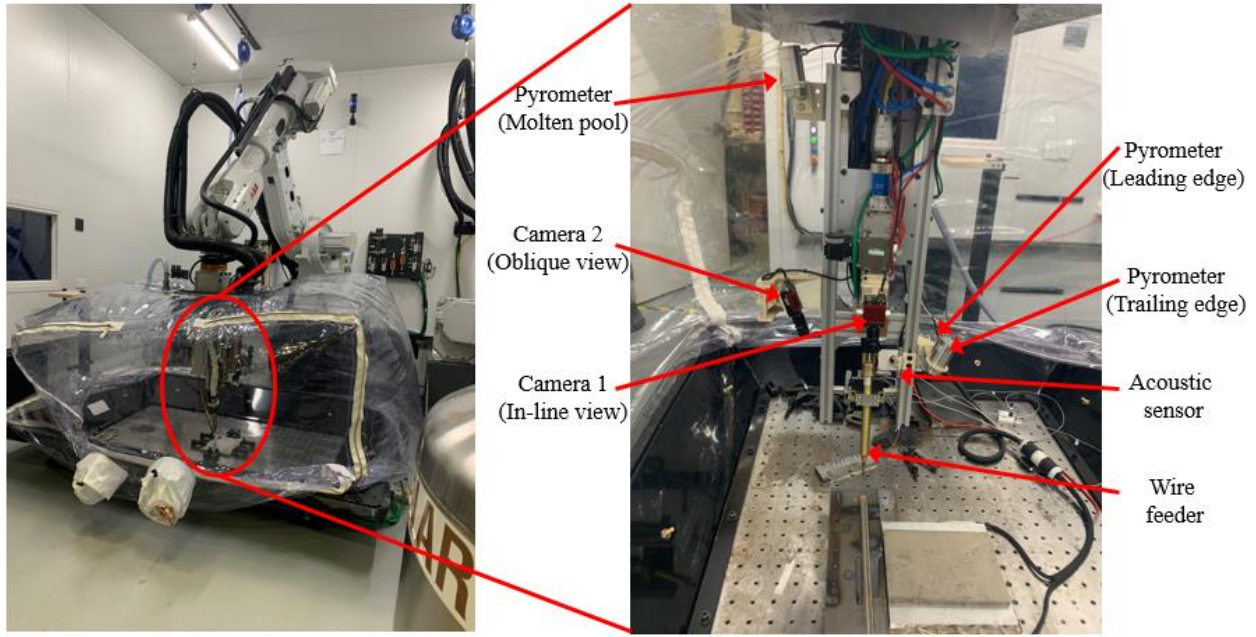

Figure 1. Integrated laser hot wire feed DED system.

## 2.2. Data preparation

In this study, thirteen 100 mm experiments were conducted for different values of process parameters, i.e., laser power (LP), travel speed (TS), wire feed rate (WFR), and hot wire power (HWP) for a 100 mm single bead deposition. Table 1 shows the combination of process parameters used for data collection for the WLAM system and the corresponding measured quality parameters. The collected characterization data for single bead deposition include four geometric properties and three microstructural parameters. The geometric parameters are bead height (H), bead width (W), fusion zone depth (D), and fusion zone area (A). The measured microstructural properties are alpha lath thickness ($\alpha$), beta grain length parallel to the build ($\beta_{\parallel}$), and beta grain length perpendicular to the build ($\beta_{\perp}$). The process parameters vary in the range of 4000-6000 W, 3.5-10 mm/s, and 40-71.3 mm/s for laser power, travel speed, and wire feed rate, respectively. The single bead deposition data consists of both the steady and transient states of the molten pool. In this paper, we will focus on the steady-state molten pool data. The pixel range for the image data from the coaxial camera is enormous, and it covers a vast portion of the unwanted region. The molten pool image is preprocessed with a selected region of interest (ROI) and cropped to reduce the image size without missing any information and keeping the data within the hardware processing capability. Hence, the coaxial camera images are cropped to a size of $481 \times 566$ to be processed and trained in MATLAB. The dataset consists of 6500 images from 13 builds, containing 500 stable molten pool images from a coaxial camera and temperature data from each build. From the 13 build dataset, 12 builds are used for training the network, while the remaining unseen one build is used in the testing phase.

Table 1. Process parameter values and measured quality parameters for the collected image and temperature data.

| Experiment # | Process parameters | | | | Quality parameters | | | | | | |
|---|---|---|---|---|---|---|---|---|---|---|---|
| | LP (W) | TS (mm/s) | WFR (mm/s) | HWP (W) | H (mm) | W (mm) | D (mm) | A (mm$^2$) | α (μm) | $β_∥$ (μm) | $β_⊥$ (μm) |
| 1 | 6000 | 6.6 | 60 | 300 | 3.46 | 11.15 | 2.05 | 37.02 | 1.38 | 424.54 | 1056.91 |
| 2 | 6000 | 5 | 60.1 | 300 | 3.81 | 11.59 | 2.16 | 42.30 | 1.39 | 466.10 | 1122.01 |
| 3 | 6000 | 3.5 | 71.3 | 400 | 4.57 | 11.83 | 2.30 | 51.16 | 1.42 | 525.88 | 1198.34 |
| 4 | 6000 | 3.5 | 50.1 | 200 | 3.72 | 12.06 | 2.20 | 43.22 | 1.38 | 485.04 | 1147.22 |
| 5 | 4500 | 5 | 48.4 | 300 | 4.07 | 9.68 | 1.92 | 38.25 | 1.32 | 452.36 | 1173.63 |
| 6 | 4500 | 5 | 50.8 | 300 | 4.17 | 9.66 | 1.93 | 39.04 | 1.33 | 457.77 | 1186.98 |
| 7 | 4500 | 10 | 50.8 | 300 | 2.93 | 8.37 | 1.51 | 24.43 | 1.29 | 321.69 | 965.91 |
| 8 | 4500 | 5 | 43 | 300 | 3.84 | 9.71 | 1.89 | 36.45 | 1.31 | 440.13 | 1142.59 |
| 9 | 4500 | 5 | 40 | 300 | 3.71 | 9.72 | 1.87 | 35.45 | 1.30 | 433.31 | 1124.77 |
| 10 | 5000 | 5 | 40 | 300 | 3.50 | 10.53 | 1.96 | 35.81 | 1.32 | 432.48 | 1109.28 |
| 11 | 4000 | 5 | 40 | 300 | 3.88 | 8.83 | 1.75 | 34.64 | 1.28 | 433.02 | 1131.84 |
| 12 | 4500 | 6.5 | 40 | 300 | 3.36 | 9.35 | 1.76 | 31.41 | 1.29 | 393.11 | 1062.43 |
| 13 | 4500 | 3.5 | 40 | 300 | 4.04 | 10.07 | 1.97 | 39.34 | 1.31 | 473.16 | 1178.81 |

The characteristics of the printed bead using the WLAM setup are quantified subsequently. No heat treatment is performed on any of the bead samples produced in this study. The printed bead is incised to identify the geometrical properties of the bead by cross-sectioning it. These specimens were first polished with SiC papers on a Struers LaboForce 100 machine. The bead geometrical properties calculation such as bead height, bead width, fusion zone depth, and the area is performed on the Keyence VHX-5000 optical microscope. The dimensional and microstructural information for a single bead deposition from cross-section is exhibited in Figure 2. The specimens are finely polished for microstructures analysis, and SEM first examined the alpha lath microstructures in the backscattered electron mode. The samples are then etched for 40–60 s in Kroll reagent (1 ml HF+2 ml HNO$_3$+50 ml H$_2$O) to emerge a flesh condition for observation. The optical microscopy under polarized light is used to distinguish the prior-β grains by their crystal orientation. The microstructural image of α lath and prior-β grains are analyzed using ImageJ software to obtain statistical information. Alpha lath thickness, beta grain length parallel to the build, and beta grain perpendicular to the bild direction are calculated for microstructural analysis.

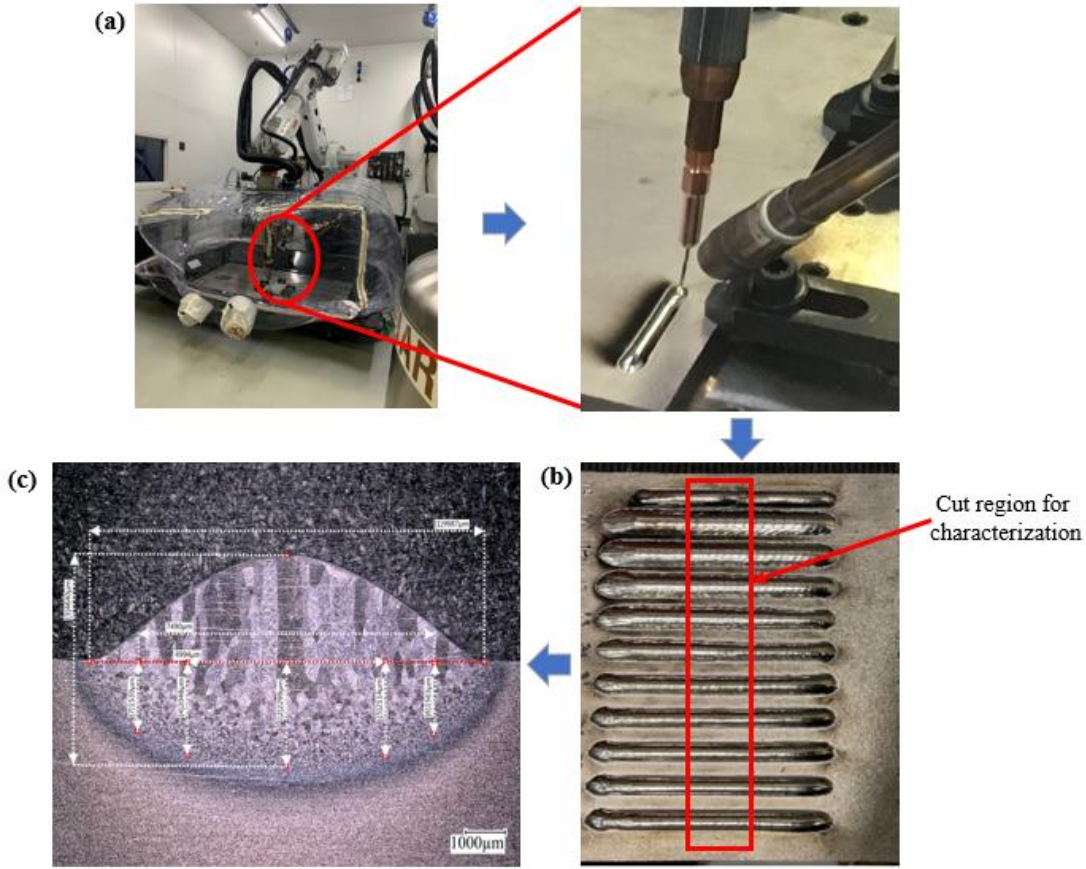

Figure 2. The experiments data collection from (a) WLAM system under a set of controlled process parameters, (b) Printed bead for different settings of process parameters (c) Characterization analysis for bead geometry and microstructure.

The MP dimensional information is extracted from the video by performing image by image analysis. The flowchart for the width and length measurement is presented in Figure 3(a). The images from the camera are accessed frame by frame to extract the red plane as a reference. The extracted red image plane is applied with thresholding for removing noisy measurement data. After thresholding, the next step is edge extraction, which is based upon the threshold applied and filter for detecting the edges. There are five basic types of filters, i.e., Prewitt, differentiation, Robert, Sobel, gradient, each suitable for a specific application [35]. Robert filter was applied to get the most accurate representation of the molten pool. Once the edges are extracted, we then define the region of interest for width and length measurement. Figure 3(b) shows the extracted width and length plot in mm for one single bead deposition experiment. The extracted width and length are used along with the MP temperature to predict the characterization data using the regression model used as a baseline.

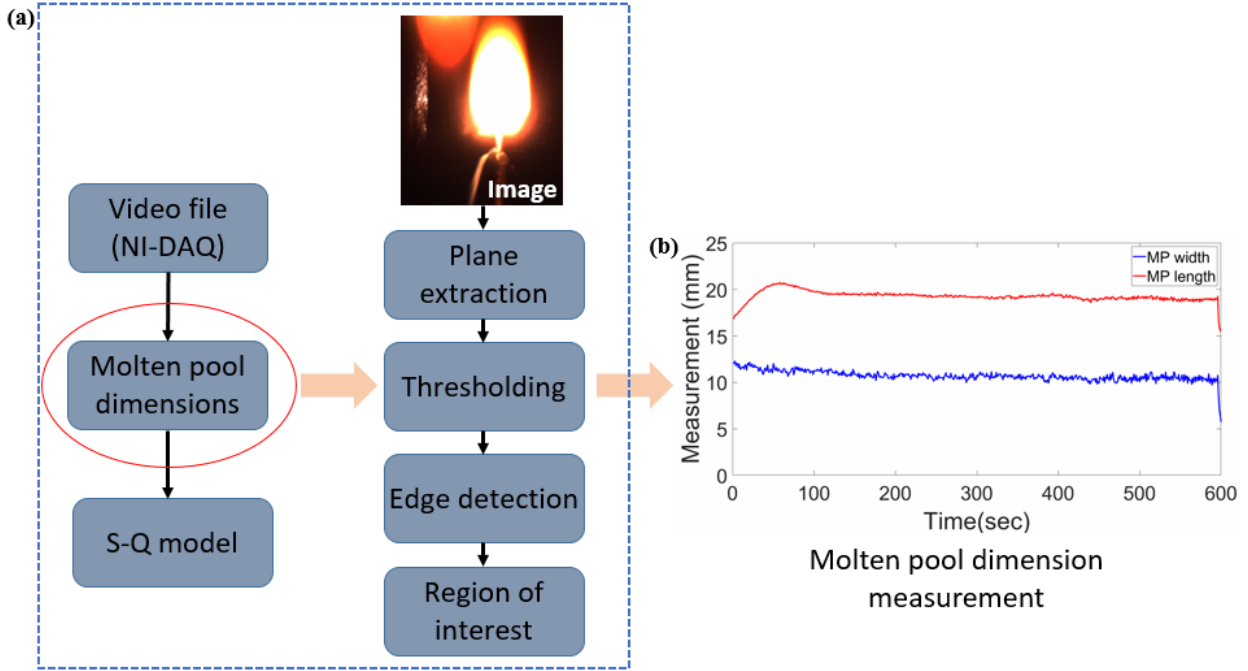

Figure 3: Image processing for molten pool dimension extraction (a) Flowchart for dimension measurement (b) Molten pool width and length plot for single bead deposition.

*2.3. Feature engineering process*

By increasing the number of features, feature engineering increases the problem's dimensionality, leading to the "curse of dimensionality" [36, 37]. Hence, before final tuning and verification of the machine learning models, it is desirable to analyze redundant or insignificant inputs. In such a case, the number of inputs may be reduced, reducing the functional complexity of the machine learning models. Feature selection is an essential technique for improving the network's performance by finding the most meaningful input features for the output prediction. It helps in improving the accuracy caused by the redundant and enormous amount of available experimental data. It is not necessary that the addition of more features into the network always lead to a better outcome. The real-world environmental data could be noisy, redundant, and invalid or have disturbances caused by the process or environment. The data used for building the model might not always be able to accurate to recognize any real pattern between the mappings. Furthermore, the corrupted data could increase network complexity in terms of time and storage. Hence, it is necessary to analyze feature pairs correlation before training the CNN model. Features correlation and redundancy were evaluated using Pearson correlation [38]. The Scikit-learn python implementation of these algorithms was used [39]. Correspondingly, Pearson correlation between feature pairs $r_{x_{ij}}$ or feature and property $r_{xy}$ uses the standard definition,

$$r_{xy} = \frac{\sum_{i=1}^{n}(x_i-\bar{x})(y_i-\bar{y})}{\sqrt{\sum_{i=1}^{n}(x_i-\bar{x})^2}\sqrt{\sum_{i=1}^{n}(y_i-\bar{y})^2}} \qquad (1)$$

Where, $n$ is the sample size, $x_i$ and $y_i$ are the individual sample points, and $\bar{x}$ and $\bar{y}$ are the sample means.

## 2.4. Sensing-quality modeling

### 2.4.1. CNN modeling

Because of the short laser-material interaction times and highly localized heat input, the thermal gradients and rapid solidification rates lead to a build-up of thermal stresses and non-equilibrium phases. Non-optimal process parameters can cause molten pool instabilities, which leads to defects and geometric distortions. Further, it has been shown that AM-specific microstructures from fast solidification of molten pool significantly affect the mechanical properties of the part, such as strength, ductility, and fatigue behavior. The molten pool behavior should be monitored and controlled carefully to obtain printed parts with designed geometry and microstructure.

Figure 4 shows the sensing-quality (S-Q) relations model, employing a convolutional neural network that takes in the sensing data from the CMOS camera and pyrometer as inputs and outputs the corresponding quality parameters based upon the sensor data. Firstly, data was collected from a WLAM system installed with sensors to record the molten pool evolution during printing. The effects of quality parameters on molten pool features and their correlations are analyzed. Next, the sensing image data collected during the build are paired with corresponding molten pool temperature data points as shown in the input block. Third, CNN architecture automatically performs feature extraction without human intervention via convolution and pooling layers to obtain meaningful features from the molten pool images. Finally, the fully connected layer takes input extracted features in conjunction with the molten pool temperature for characterization parameter estimation, i.e., the geometric and microstructural properties. Two networks are trained on the data collected on a WLAM system under a set of controlled process parameters, one for estimating geometric parameters and the other for the microstructural parameters. Once the networks are trained, they can predict the quality parameters directly from unseen real-time sensor data.

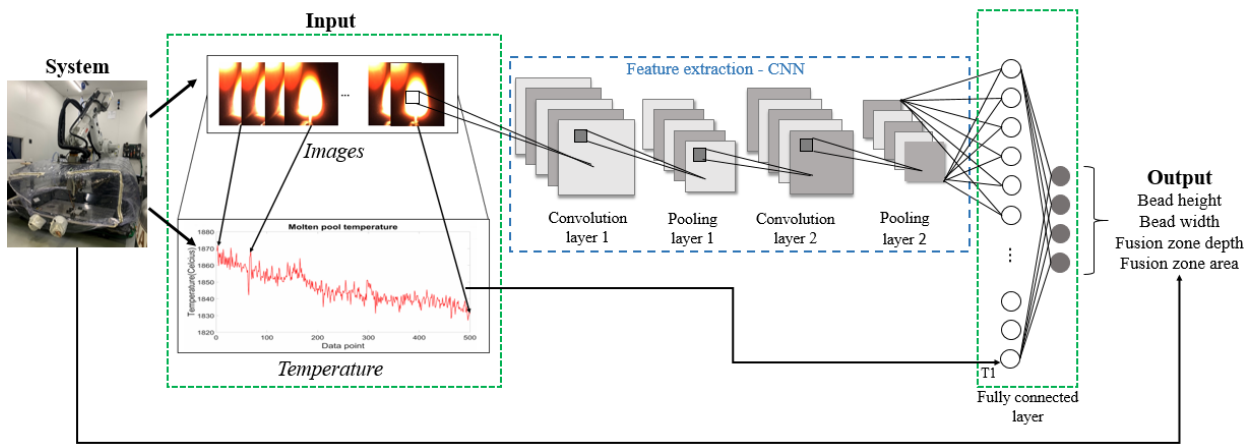

Figure 4. Sensing-quality relations modeling using CNN for quality parameter prediction using molten pool images and temperature data.

Specifically, a convolutional neural network is based on an artificial neural network, a popular machine learning method inspired by neurons in the human body. In recent years, a plethora of work has been done using CNN on experimental data for various applications [40-43]. CNN is widely used in image processing because it learns local spatial relations by exploiting the features in a hierarchical nature. CNN takes advantage of the multi-dimensional image input, shares weights along with the layers, and knows the intricate features concerning each other from high-dimensional image data to output classes or values. CNN consists of different layers such as a convolutional layer, activation function, pooling layer, fully connected layer, batch normalization, dropout, etc., for classification and regression tasks. The input image is first processed by a set of kernels, also known as filters. The convolutional layer is applied to extract features by convolving the features from the previous layer and shifting the filter as specified by the stride. The filters are learnable weights that are continuously updated during learning, while the size of the filter is user-defined. The feature map F is represented in Eq. (2) as,

$$F(i,j,k) = \sum_{m=0}^{M-1} \sum_{n=0}^{N-1} \sum_{l=0}^{L-1} E(i+m, j+n, l) K_k(m.n, l) + b_k \qquad (2)$$

Where, $E$ is the extracted features from the previous layer, $K_k$ is the applied filter, $b_k$ is the bias, (M, N, L) is the size of the filter, and L is the 3rd dimension of the previous feature map layer.

The feature map represents the input, i.e., the molten pool image response to the feature defined by the input filter. With the addition of more convolutional layers, the features become more abstract and directed towards the output prediction/classification problem. After the convolution layer, a non-linearity activation function is applied to the network to introduce non-linearity. The standard activation functions that are applied are ReLU, Tanh, Sigmoid, LeakyReLU, etc. In this work, the hyperbolic tangent (tanh) activation function is used. A batch normalization layer is applied to address the issue of covariant shift and non-uniform scaling ranges in the internal layers. This regularization layer reduces overfitting and network malfunction due to improper learning rate and speeding up the training process [44]. Pooling layers are added to downsample the data dimension and introduce local invariance, making the feature map less sensitive to locations. Standard pooling functions are global pooling, average pooling, max pooling, etc. The pooling layer decreases the computational volume of data and increases the robustness of the algorithm. The last pooling layer's output is an input feature vector to a fully connected layer to predict the final process parameters. The output of the fully connected layer is given by Eq. (3) as,

$$X_{i+1} = W_i . X_i + b_i \qquad (3)$$

Where, $X_i$ is the i$^{th}$ fully connected layer, $W_i$ is the weight matrix, and $b_i$ is the bias vector.

The different layers are connected in a user-defined manner to form a network that is trained using backward propagation that uses a gradient descent technique. The learning parameters are adjusted during the training to achieve accurate prediction of process parameters and minimize the training loss function, i.e., mean square error in our case as given in Eq. (4),

$$Loss_{MSE} = \sum_{1}^{N} l(y^n, \hat{y}^n)/N \qquad (4)$$

Where, $l$ is the loss function, $y^n$ is the target value, $\hat{y}^n$ is the predicted value, and N is the size of the training data.

*2.4.2. Regression modeling*

Different regressions techniques such as linear regression, Gaussian process regression (GPR), polynomial regression, and support vector regression (SVR) are popular function estimation methods that can be used for prediction modeling [45]. Each method is suitable for a specific application based upon the dataset and modeling approach. The regression based model is trained using a support vector machine for geometrical properties modeling, while Gaussian process regression is utilized for microstructural properties prediction.

SVR works by minimizing the generalized error bound instead of minimizing the observed training error so as to achieve generalized performance. The generalization error bound is the combination of the training error and a regularization term that controls the hypothesis space's complexity. The goal is to search for a function that predicts the output based upon the prediction error between the actual and predicted value [46]. The kernel function is used to fit the input features, while the cost function works by searching the optimal weight parameters for minimizing the total accumulated error. With no prior knowledge about the complex relationship between the molten pool data and geometric properties, a non-linear SVR with linear kernel function is used for mapping the input to the output.

The Gaussian process regression is another ML technique utilized to map the complex non-linear relation of the sensing data directly to the post-processing microstructural properties. GPR model is shown effective in modeling compositions-process-property correlations [47]. It is advantageous because of its adaptability to a small dataset and provides uncertainty measurement on the estimated value. Gaussian process regression is a probabilistic nonparametric approach based upon the kernel function. GPR operates by estimating the hyperparameters for the covariance function, the noise variance, and the coefficient of the basis function while the model is being trained. The hyperparameters of the covariance functions are tuned by maximizing the log marginal likelihood of the input data during training. The initial values for the choice of covariance function and basis function can be specified by the user. A squared exponential kernel with a separate length scale for each predictor is used as the covariance function to deal with complex input-output relations. A fully independent conditional approximation and a subset of regressors approximation are used as predict method and fit method based upon the nature of the available dataset. Matlab version 2020b was used to implement CNN, SVR, and GPR models.

*2.5. Model evaluation*

In order to predict the process parameters accurately, the network structure, input, and output need to be evaluated along with the structure generalization capability. When constructing a CNN, the more input layers result in a higher quality of the extracted features. However, this is not always the case because of dissipating or shooting gradient or degenerative models when using deeper networks. When training a deep NN, the gradient calculated using backpropagation is responsible for updating the network's weights. Poorly designed network structure causes the gradient to shoot or vanish, resulting in a slower model learning process and lower accuracy. Hence, the CNN architecture and the layers are optimized to suit the specific application for designed input and output. The multi-modality CNNs are trained to predict the geometric and microstructural parameters using two separate models.

Generally, the developed model returns a perfect result under specific training and testing samples. However, its performance could be worse if the training and testing samples do not represent the entire dataset, even if the samples belong to the same dataset. Cross-validation is a common way to avoid this problem. We implemented 6-fold cross-validation where the dataset is split into a 1:5 ratio. The smaller set is used for testing, and the larger set is used for training, i.e., 5-parts training and 1-part validation. The generalization is gauged by performing 6-fold cross-validation where is the network is trained on 6000 molten pool image and temperature data samples and tested on the 500 unseen samples.

The evaluation criteria used for performance comparison are the root mean square error (RMSE), standard deviation (SD) for RMSE, and relative percentage error (RE) between the actual and predicted parameters, given in Eqs. (5, 6) respectively. $y_i$ is the measured, and $y_i'$ is the model predicted process parameter. Normalized RMSE (NRME) is another error evaluation metric for different scales of process parameters, as given in Eq. (7). The RMSE is normalized to 1.6483 mm for bead height, 3.6976 mm for bead width, 0.7970 mm for fusion zone depth, and 26.7311 mm$^2$ for fusion zone area. The normalization values used for microstructural properties prediction are 0.1423 µm, 204.1858 µm, 232.4281 µm for alpha lath thickness, beta grain length parallel to the build, and beta grain length perpendicular to the build, respectively.

$$RMSE = \sqrt{\frac{1}{n}\sum_{i=1}^{n}(y_i' - y_i)^2} \tag{5}$$

$$RE = \frac{1}{n}\sum_{i=1}^{n}\left|\frac{y_i' - y_i}{y_i}\right| \times 100\% \tag{6}$$

$$NRMSE = \frac{RMSE}{(y_{max} - y_{min})} \times 100\% \tag{7}$$

Multi-modality CNN architectures for geometry and microstructure are trained using the mean square error as a loss function between the actual and predicted control parameters for the specific build. The evaluation metric for validating the training and testing is the convergence and the accuracy performance of the trained model. The convergence is achieved by calculating the gradient of the loss function and propagating it back through the network. The network weight incoming from the neurons is altered to minimize the loss function during backpropagation. The hyperparameters influencing the backpropagation and controlling the network convergence are the learning rate, momentum, and velocity. The optimization parameters utilized for training the geometric model (GM) and microstructural model (MM) are detailed in Table 2.

Table 2. Optimized hyperparameter values for the CNN architectures.

| Hyper-parameter | CNN model | |
|---|---|---|
| | Geometric model (GM) | Microstructural model (MM) |
| Learning rate | 10$^{-5}$ | 10$^{-5}$ |
| Momentum | 0.9 | 0.9 |
| Epoch | 10 | 10 |
| Batch size | 10 | 10 |

# 3. Results

## 3.1. Feature analysis and selection

In this study, the sensing data includes the camera images and temperature measurement of the molten pool (MP), leading edge (LE), and trailing edge (TE) of the molten pool region. Table 3 shows the Pearson correlation factors between the measurable sensing parameters and the post-processing characterization data. Molten pool dimensional features are extracted from the image data; the width and length of the pool as described in section 4.3. Analyzing the Pearson correlation between the measurable data and quality parameters identified the highest correlation between the molten pool temperature and the characterization data. Hence, along with image data, molten pool temperature is used as input to the CNN based S-Q model.

CNN models directly use the camera images with minimal preprocessing as input to the networks along with the MP temperature. The first CNN network is trained to output four geometric parameters, and the other network is used for predicting the three microstructural data. The development of two separate models for predicting geometric and microstructural parameters allows for getting low-performance error and improved accuracy.

Table 3. Pearson correlation matrix between the sensing data and quality parameters.

| Variable | Sensing data | | | | | Characterization parameters | | | | | | |
|---|---|---|---|---|---|---|---|---|---|---|---|---|
| | MP width (mm) | MP length (mm) | MP temp (°C) | TE Temp (°C) | LE Temp (°C) | H (mm) | W (mm) | D (mm) | A (mm$^2$) | $\alpha$ (µm) | $\beta_\parallel$ (µm) | $\beta_\perp$ (µm) |
| MP width (mm) | 1 | 0.98 | 0.88 | 0.25 | 0.94 | 0.24 | 0.90 | 0.81 | 0.69 | 0.90 | 0.55 | 0.15 |
| MP length (mm) | 0.98 | 1 | 0.90 | 0.29 | 0.91 | 0.16 | 0.89 | 0.78 | 0.63 | 0.91 | 0.48 | 0.08 |
| MP temp (°C) | 0.88 | 0.90 | 1 | 0.40 | 0.76 | 0.08 | 0.85 | 0.73 | 0.55 | 0.88 | 0.38 | 0.01 |
| TE temp (°C) | 0.25 | 0.29 | 0.40 | 1 | 0.16 | -0.13 | 0.42 | 0.32 | 0.14 | 0.31 | 0.11 | -0.04 |
| LE temp (°C) | 0.94 | 0.91 | 0.76 | 0.16 | 1 | 0.20 | 0.83 | 0.73 | 0.63 | 0.78 | 0.51 | 0.14 |
| H (mm) | 0.24 | 0.16 | 0.08 | -0.13 | 0.20 | 1 | 0.39 | 0.64 | 0.84 | 0.43 | 0.89 | 0.94 |
| W (mm) | 0.90 | 0.89 | 0.85 | 0.42 | 0.83 | 0.39 | 1 | 0.95 | 0.82 | 0.91 | 0.73 | 0.39 |
| D (mm) | 0.81 | 0.78 | 0.73 | 0.32 | 0.73 | 0.64 | 0.95 | 1 | 0.95 | 0.88 | 0.89 | 0.63 |
| A (mm$^2$) | 0.69 | 0.63 | 0.55 | 0.14 | 0.63 | 0.84 | 0.82 | 0.95 | 1 | 0.80 | 0.97 | 0.79 |
| $\alpha$ (µm) | 0.90 | 0.91 | 0.88 | 0.31 | 0.78 | 0.43 | 0.91 | 0.88 | 0.80 | 1 | 0.64 | 0.30 |
| $\beta_\parallel$ (µm) | 0.55 | 0.48 | 0.38 | 0.11 | 0.51 | 0.89 | 0.73 | 0.89 | 0.97 | 0.64 | 1 | 0.90 |
| $\beta_\perp$ (µm) | 0.15 | 0.08 | 0.01 | -0.04 | 0.14 | 0.94 | 0.39 | 0.63 | 0.79 | 0.30 | 0.90 | 1 |

## 3.2. CNN model performance for geometry parameters

The multi-modality CNN architecture for predicting the geometric parameters using the image and MP temperature data consists of a total of 24 layers, where the first 21 layers are used for image feature extraction. The remaining three layers are for predicting the final control parameters along with the temperature data feature. The network uses batch normalization, clipped ReLU activation function, and a dropout of 50%. The multi-modality CNN consists of four convolution layers interlaced with two global average pooling layers. Following the second global average pooling, the resulting output is unrolled into a vector and feed into a fully connected layer of dimension 100, which is reduced to a size of 3 before concatenating with the temperature feature.

The final layer consists of 4 nodes, based upon the four geometric parameters' prediction. The geometric model (GM) is trained for an epoch of 10 using 6000 training samples and 500 unseen test samples of image and temperature data collected during 13 builds for the molten pool condition. The CNN based GM uses 6-fold cross-validation and the results discussed below are for the unseen test dataset.

Table 4 shows the performance of the CNN model predicting the four final build geometric parameters; H, W, D, and A. The RMSE for height, width, depth, and the area is 0.30 mm, 0.50 mm, 0.09 mm, and 3.39 mm$^2$. The RE for all the four parameters' prediction is less than 10 %, which is 8.04 % for height, 4.53% for width, 4.14 % for depth, and 8.74 % for the area. The NRMSE value is 18.35%, 13.62 %, 10.84 %, and 12.70 % for height, width, depth and area respectively. The SD value for RMSE using 6-fold CV is 0.11, 0.24, 0.05, and 1.90 for height, width, depth, and area, respectively.

The NRMSE is highest for bead height as it is shown to have the least correlation with the MP temperature. The RE is relatively higher for bead's geometric shape height and fusion zone area; this is caused by their lower correlations with regard to the MP geometric and thermal profiles, as indicated by Table 3. Molten pool surface area is found to be a poor indicator of cooling rate at different locations in the molten pool. Therefore, it cannot be relied upon to achieve targeted part bead geometry control [48]. A more detailed characterization of molten pool dynamics and subsurface information is needed to achieve the ultimate goal of in situ quality control of printed parts [49].

Table 4. Prediction performance of multi-modality CNN model for build geometric parameters using RMSE, RE, NRMSE, and SD.

| Error metric | CNN model | | | |
| --- | --- | --- | --- | --- |
| | Geometric parameters | | | |
| | H (mm) | W (mm) | D (mm) | A (mm$^2$) |
| RMSE | 0.30 | 0.50 | 0.09 | 3.39 |
| RE | 8.04 % | 4.53 % | 4.14 % | 8.74 % |
| NRMSE | 18.35 % | 13.62 % | 10.84 % | 12.70 % |
| SD | 0.11 | 0.24 | 0.05 | 1.90 |

*3.3. CNN model performance for microstructural parameters*

The multi-modality CNN architecture for predicting the microstructural parameters using the image and MP temperature data consists of a total of 24 layers same as the geometric model. The unrolled output from the global average pooling layer is fed into a fully connected layer of dimension 110, which is reduced to a size of 3 before concatenating with the temperature feature. The final prediction layer consists of three nodes based upon the three microstructure parameters' prediction. The microstructure model (MM) is trained using 6000 training samples and 500 unseen test samples of MP image and temperature data collected during 13 builds. The CNN based MM uses 6-fold cross-validation, and the results discussed below are for the unseen test dataset.

Table 5 shows the prediction performance of the CNN model for the microstructural properties; alpha lath thickness($\alpha$), beta grain length parallel to the build direction ($\beta\parallel$), and beta grain length perpendicular to the build ($\beta\perp$). The RMSE is 0.02 µm for $\alpha$, 20.27 µm for $\beta\parallel$, and 30.26 µm for $\beta\perp$. The RE for $\alpha$, $\beta\parallel$, and $\beta\perp$ is 1.15 %, 5.84 %, and 3.73 %, respectively. The value for normalized RMSE is 12.31 %, 9.92 %, and 13.02 % for α, β∥, and β⊥ , respectively. The SD for the 6-fold cross-validation is 0.01 for α, 18.00 for β∥, and 31.05 for β⊥.

As seen from Table 5, the lowest RMSE, RE, and SD are exhibited for α phase thickness as described by strong Pearson correlation with sensing data from Table 3. The prediction error for β grain length in parallel and perpendicular is relatively larger. The reason could be the data collection methodology, where the length of β length is relatively difficult to accurately measure due to the end of *β* grain mixed with finer equiaxed *α/β* microstructures.

Table 5. Prediction performance of multi-modality CNN model for build microstructure parameters using RMSE, RE, NRMSE, and SD.

| Error metric | CNN model | | |
| --- | --- | --- | --- |
| | Microstructural parameters | | |
| | A (μm) | β∥ (μm) | β⊥ (μm) |
| RMSE | 0.02 | 20.27 | 30.26 |
| RE | 1.15 % | 5.84 % | 3.73 % |
| NRMSE | 12.31 % | 9.92 % | 13.02 % |
| SD | 0.01 | 18.00 | 31.05 |

*3.4. CNN and regression model comparison*

The performance of the CNN model is compared with another ML regression model trained on the same dataset. The regression model is used as a baseline for comparison with the CNN model's prediction. Figure 5 shows the general framework that employs the sensing-quality regression model for property prediction using molten pool dimensional information and temperature data. The input to the regression model is the molten pool dimensional information; width, length, and MP temperature data. The MP temperature is used as input as it is shown to have the highest correlation with the characterization properties as described in section 3.2. The model's output is the geometrical and microstructural properties as that of CNN models. Two regression models are trained, one predicting the four geometric parameters and the other for predicting three microstructural data.

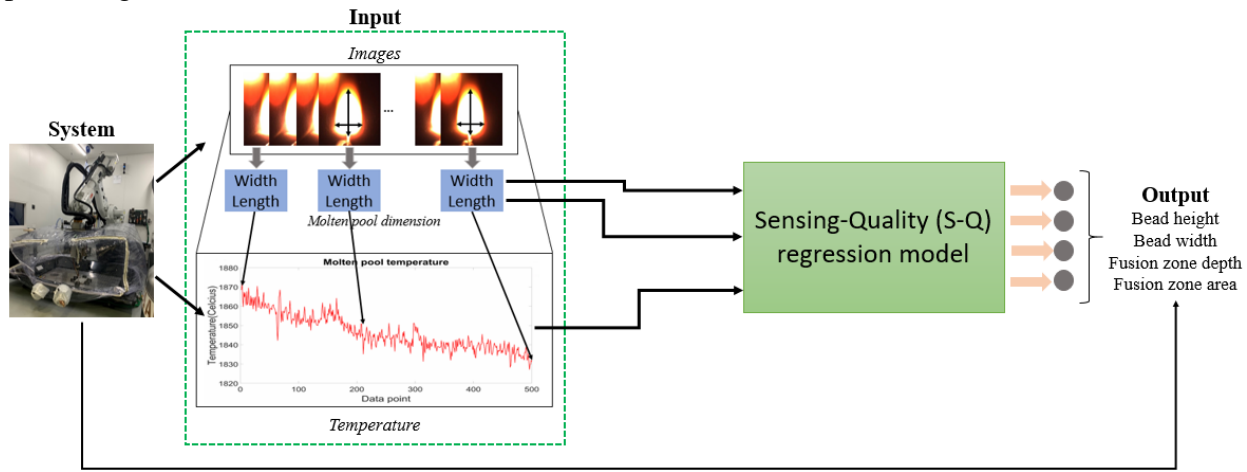

Figure 5. Sensing-quality regression model for quality parameter prediction using molten pool dimensional information and MP temperature data.

The comparison between CNN and regression model for geometric and microstructural parameter prediction is presented in Table 6. The geometric model's comparison shows that the RMSE for the CNN model is overall better than the regression model. The bead width performance is considerably lower for the regression model since molten pool width is used as model input, which is strongly correlated to the printed bead width. The prediction error of fusion zone depth is lower for the regression model. It may result from a relatively stronger coupling effect of fusion zone depth with the molten pool surface length and width than bead height and fusion zone area. Prediction error for fusion zone area is lower for the CNN model as the molten pool image data provide more information for area prediction. The SD for all the geometrical parameter predictions is lower for the CNN model showing that the prediction is clustered around the mean, making the model reliable.

Microstructural properties prediction comparison shows that the model trained using CNN is better for all the error measures. Features extracted by the CNN directly from the MP image data contain more relevant information for microstructural properties prediction. The extracted image features and temperature data provide intricate details helpful for in situ microstructural properties prediction.

Table 6. Comparison between regression and multi-modality CNN model for predicting the characterization parameters using RMSE, RE, NRMSE, and SD.

| Error metric | Data-driven model | | | | | | | | | | | | | |
|---|---|---|---|---|---|---|---|---|---|---|---|---|---|---|
| | CNN model | | | | | | | Regression model | | | | | | |
| | GM | | | | MM | | | GM | | | | MM | | |
| | H (mm) | W (mm) | D (mm) | A (mm²) | α (μm) | $β_∥$ (μm) | $β_⊥$ (μm) | H (mm) | W (mm) | D (mm) | A (mm²) | α (μm) | $β_∥$ (μm) | $β_⊥$ (μm) |
| RMSE | 0.30 | 0.50 | 0.09 | 3.39 | 0.02 | 20.27 | 30.26 | 0.36 | 0.35 | 0.06 | 7.94 | 0.03 | 29.51 | 53.03 |
| RE | 8.04% | 4.53% | 4.14% | 8.74% | 1.15% | 5.84% | 3.73% | 8.73% | 3.49% | 3.28% | 18.72% | 1.98% | 6.86% | 5.02% |
| NRMSE | 18.35% | 13.62% | 10.84% | 12.70% | 12.31% | 9.92% | 13.02% | 21.66% | 9.48% | 7.80% | 29.70% | 21.63% | 14.45% | 22.82% |
| SD | 0.11 | 0.24 | 0.05 | 1.90 | 0.01 | 18.00 | 31.05 | 0.32 | 0.26 | 0.06 | 6.98 | 0.01 | 11.36 | 19.37 |

## 4. Discussions

*4.1. Property variations against molten pool shape and temperature*

The WLAM is a high-dimensional complex process involving many interrelated parameters. In this paper, a sensing system using one CMOS camera and three infrared pyrometers has been developed for the WLAM process. Real-time tracking and measurement are achieved during the printing processes, and post-process characterization is performed on the printed single bead deposition. As described in section 2.2, only the steady-state molten pool sensing data from the 13 build samples are used to train and test the network. The molten pool images are paired with corresponding temperature profile data from the pyrometers to be used for sensing-quality relations modeling. The involved set process parameters are responsible for generating the corresponding molten pool sensed data. Understanding the process parameters and their impact on the molten pool properties is paramount for determining the post-processing characterization data. The geometric and microstructural properties of the final build are one of the indicators for determining the quality of the structure. The utilization of controlled process parameters leads to a stable molten pool behavior by keeping the thermal history, residual stress, energy input to a constant value. Thus, controlling the process parameters for the WLAM system based upon the molten pool behavior will allow for monitoring the characterization parameters.

Table 1 shows the variation of characterization parameters, both geometric and microstructural properties, against the process parameters. Figure 6 shows the effect of change in the process parameters on the molten pool behavior i.e., molten pool image and temperature measurement in the steady-state region. Figure 6(a) presented how the increase in LP causes the MP to grow and temperature to increase. As seen from Figure 6(b), the change in TS shows that the MP shrinks and temperature decreases as TS go up. Figure 6(c) describes the change in MP behavior based upon the change in WFR. The varying sensing data can directly be correlated to the characterization properties listed in Table 1 and allows for building a quality estimation and monitoring framework.

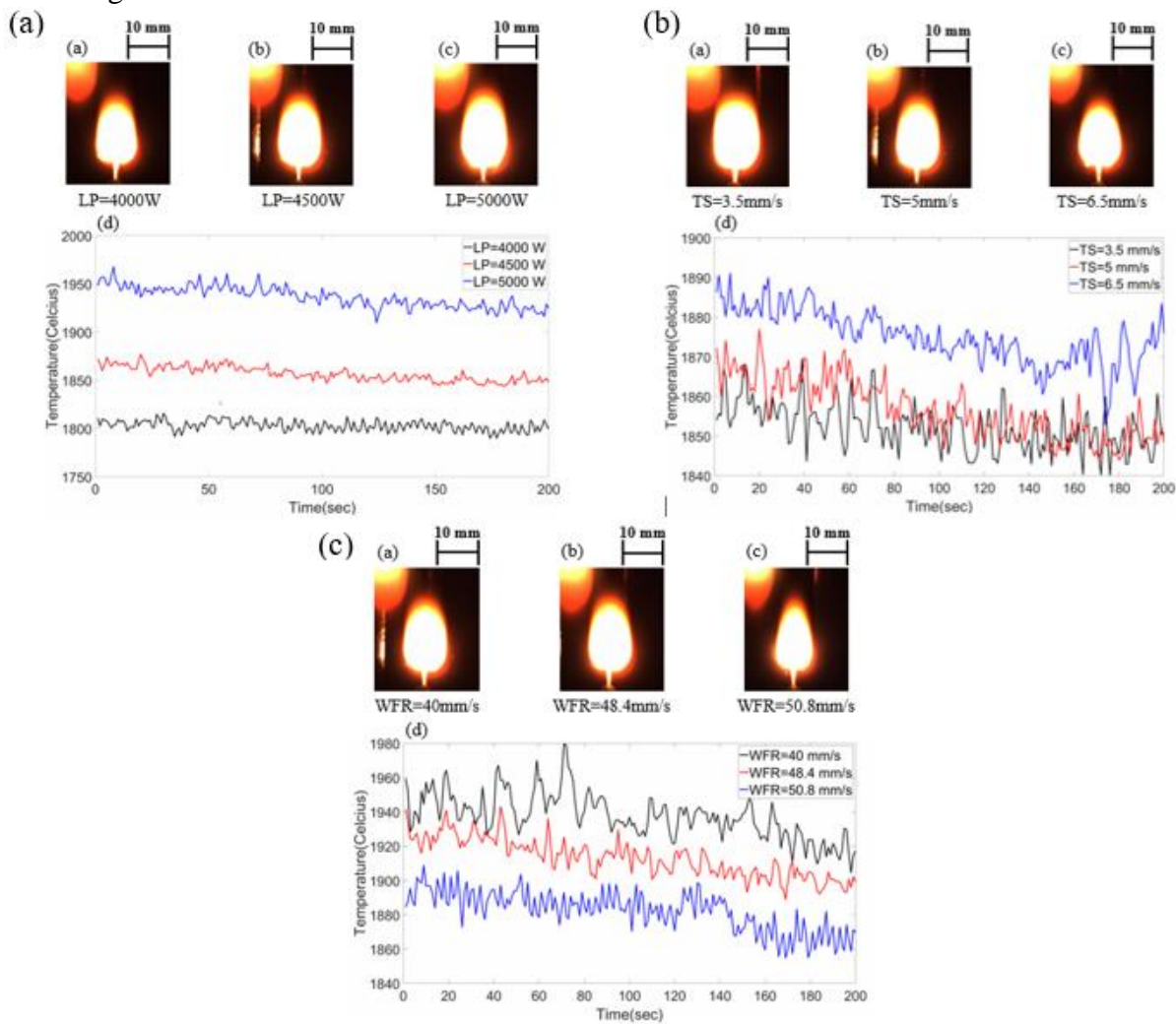

Figure 6. Illustration of samples for stable molten pool image and the corresponding molten pool temperature data for change in (a) laser power at a fixed travel speed of 5 mm/s and wire feed rate of 40 mm/s, (b) travel speed at a fixed laser power of 4500 W and wire feed rate of 40 mm/s, (c) wire feed rate at a fixed laser power of 4500 W and travel speed of 5 mm/s.

## 4.2. Influence of temperature on the multi-modality CNN models performance

Raghavan et al. [50] developed a heat transfer and fluid flow model for the laser DED process to examine how changes in processing parameters affect the relationships between molten pool surface, thermal cycles, and solidification parameters in Ti-6Al-4V. Molten pool surface area is found to be a poor indicator of cooling rate at different locations in the molten pool. Therefore, it cannot be relied upon to achieve targeted part quality control. It is shown that monitoring or controlling only the molten pool surface area can be insufficient for achieving targeted microstructures and mechanical properties. The geometric and microstructural accuracy of the printed part depends on the coupling effects of bead height, width, and depth of the molten pool. The printed part's mechanical property and defect population are sensitive to the molten pool's thermal history, which is incorporated into the model using the multi-modality function.

As seen from the Pearson correlation presented in Table 3, both molten pool and leading edge temperature are highly correlated with the characterization data compared to the trailing edge temperature measured behind the molten pool. However, Figure 7 and 8 verifies that the overall performance of the multi-modality CNN models (both GM and MM) using molten pool image (I) and MP temperature data generally outperforms the other two models in terms of error metrics used for evaluation. The temperature of the molten pool at the center near the laser heating location is a more representative temperature feature for the molten pool during printing, providing more useful information for S-Q modeling.

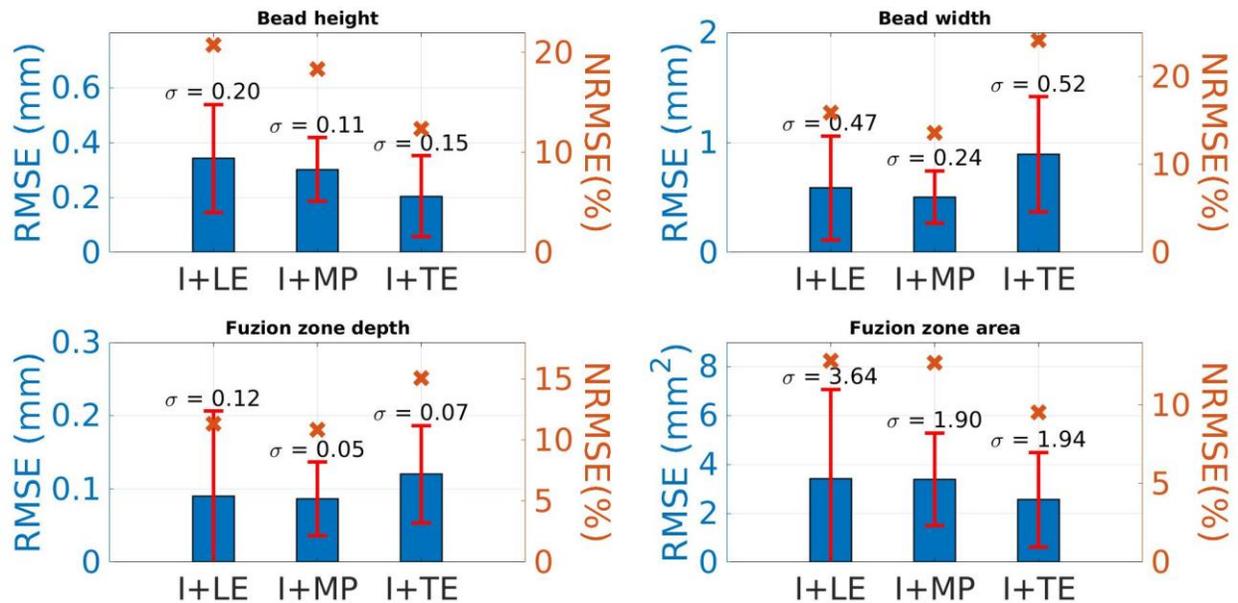

Figure 7. RMSE, NRMSE, and SD comparison for multi-modality CNN geometric model with different temperature measurements as input along with image data.

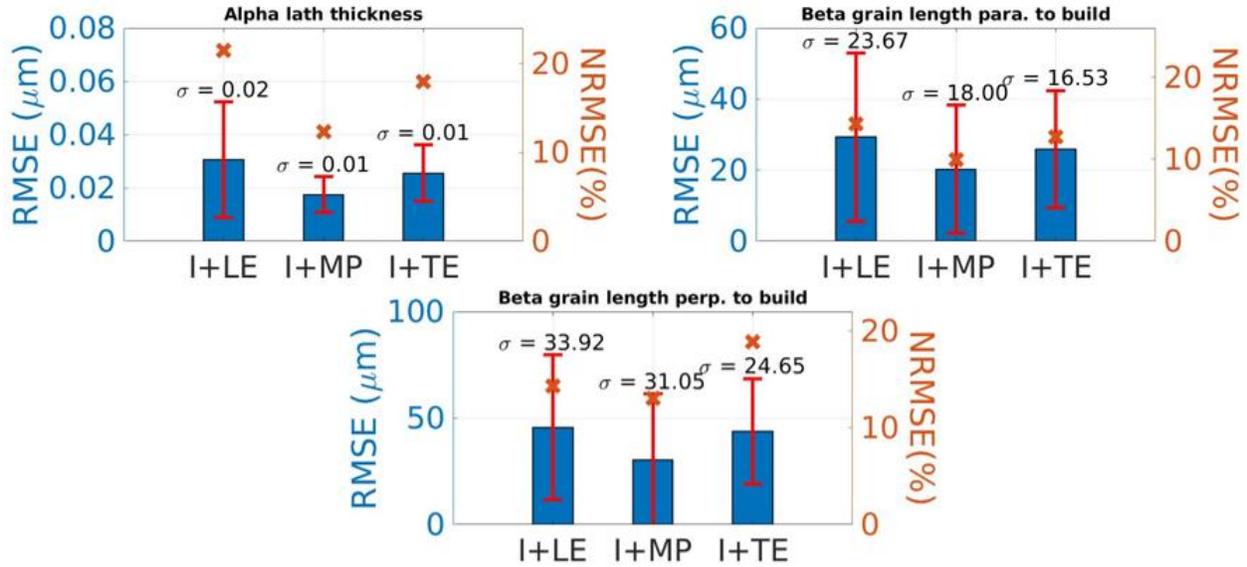

Figure 8. RMSE, NRMSE, and SD comparison for multi-modality CNN microstructural model with different temperature measurements as input along with image data.

*4.3. S-Q models for in situ quality monitoring*

The developed model can be used in real-time using the sensor data to predict the printed part geometry and microstructural properties. It serves as an in situ quality monitoring framework saving post-processing characterization cost. Most of the final printed parts are tested destructively for quality check. This results in high manufacturing costs, low volume of production, and is time-consuming. The trained S-Q model presented in Figure 4 will enable online estimation of characterization data. The developed model provides a quantitative and collaborative prediction framework for the build quality of the WLAM process. The trained geometric and microstructural model can directly take input from the WLAM sensing system. The input data is preprocessed using the data preparation method described in section 2.2. The processed data is used by the GM and MM to predict the post processing characterization data quantitatively. The overall performance accuracy for the geometric and microstructural properties prediction is less than 20%.

## 5. Conclusion

In this paper, we presented a multi-modality CNN based sensing-quality model trained to establish the correlation of the molten pool image and temperature data to the desired bead quality. Experimental sensing data is collected on a WLAM system under set controlled process parameters for single-bead deposition. Post-processing analysis is performed on the deposited bead for obtaining characterization properties. The characterization data used in this paper includes four geometric properties; bead height, bead width, fusion zone depth, and fusion zone area. The microstructural parameters used in this study are alpha lath thickness, beta grain length parallel to the build, and beta grain length perpendicular to the build. Feature analysis is performed on the sensing and characterization data to remove redundant and noisy information. The two developed models are trained to predict the geometrical and microstructural parameters separately. The performance of the CNN models is compared with the regression used as the baseline. It is seen that the prediction performance of the CNN model is overall better than the regression model. The

CNN geometric model prediction accuracy is 8.04 %, 4.53 %, 4.14 %, and 8.74 % for bead height, bead width, fusion zone depth, and fusion zone area, respectively. The microstructural model's accuracy is 1.15 % for alpha lath thickness, 5.84 % for beta grain length parallel to the build, and 3.73 % for beta grain length perpendicular to the build. The developed models allow for real-time prediction of geometric and microstructural properties of the final build directly from the sensor data. The CNN models can be used as an in situ quality assurance and control framework for monitoring and controlling the final print quality in real-time.

**Conflict of interest**
The authors do not have any conflict of interest.